 \documentclass[12pt,smallabstract,smallcaptions]{dccpaper}

\usepackage{epsfig}
\usepackage{amsmath}
\usepackage{amssymb}
\usepackage{color}
\usepackage{url}
\usepackage[utf8]{inputenc}
\usepackage{graphicx}
\usepackage{amsmath}
\usepackage{amsthm}
\newlength{\figurewidth}
\newlength{\smallfigurewidth}

\newtheorem{algorithm}{Algorithm}
\newtheorem{theorem}{Theorem}
\newtheorem{lemma}{Lemma}
\newtheorem{corollary}{Corollary}
\newtheorem{property}{Property}

\newtheorem{definition}{Definition}

\newcommand\makeset[1]{\ensuremath{\texttt{makeset}(#1)}}
\newcommand\dmerge[1]{\ensuremath{\texttt{merge}(#1)}}
\newcommand\dsplit[1]{\ensuremath{\texttt{split}(#1)}}
\newcommand\dshift[1]{\ensuremath{\texttt{shift}(#1)}}
\newcommand\occ{\ensuremath{\mathsf{occ}}}

\begin{document}

\title
{\large
\textbf{Decompressing Lempel-Ziv Compressed Text}
}

\author{Philip Bille$^{\ast}$, Mikko Berggren Ettienne$^{\ast}$, Travis Gagie$^{\dag}$,\\Inge Li Gørtz$^{\ast}$, and Nicola Prezza$^{\ddagger}$\\[0.5em]
{\small\begin{minipage}{\linewidth}\begin{center}
\begin{tabular}{ccc}
$^{\ast}$Technical University   & $^{\dag}$Dalhousie University    & $^{\ddagger}$University of Pisa\\
of Denmark                      & 6050 University Ave.              & Largo Bruno Pontecorvo 3\\
2800 Kgs.\ Lyngby               & Halifax, Canada                   & Pisa, Italy\\
Copenhagen, Denmark             & \url{travis.gagie@dal.ca}      & \url{nicola.prezza@di.unipi.it}\\
\url{{phbi,inge}@dtu.dk}\ ,     &&\\         
\url{mikkobe@gmail.com}
\end{tabular}
\end{center}\end{minipage}}
}

\maketitle
\thispagestyle{empty}

\begin{abstract}
   We consider the problem of decompressing the Lempel--Ziv 77 representation of a string $S$ of length $n$ using a working space as close as possible to the size $z$ of the input. The folklore solution for the problem runs in $O(n)$ time but requires random access to the whole decompressed text. Another folklore solution is to convert LZ77 into a grammar of size $O(z\log(n/z))$ and then stream $S$ in linear time. In this paper, we show that  $O(n)$ time and $O(z)$ working space can be achieved for constant-size alphabets. 
On general alphabets of size $\sigma$, we describe (i) a trade-off achieving $O(n\log^\delta
\sigma)$ time and $O(z\log^{1-\delta}\sigma)$ space for any $0\leq
\delta\leq 1$, and (ii) a solution achieving $O(n)$ time and
$O(z\log\log (n/z))$ space. The latter solution, in particular, dominates both folklore algorithms for the problem.
Our solutions can, more generally, extract any
specified subsequence of $S$ with little overheads on top of the linear
running time and working space. As an immediate corollary, we show that our
techniques yield improved results for pattern matching problems on
LZ77-compressed text.
\end{abstract}

\section{Introduction}

In this paper we consider the following problem: given an LZ77 representation of
a string $S$ of length $n$, decompress $S$
and output it as a stream in left-to-right order (without storing it explicitly).
Our goal is to solve this problem in as little space as possible, i.e. close to the size $z$ of the compressed input string, and as fast as possible. 
This problem is fundamental 
%and of great relevance 
in applications dominated by big repetitive data, where information has to be analyzed on-the-fly due to limitations in storage resources. 

%Moreover, this strategy cannot be used to decompress \emph{arbitrary} substrings (just the whole string).

The folklore solution for the Lempel-Ziv decompression problem achieves  linear time, but requires random access to the whole string. 
A better solution is to convert LZ77 into a straight-line program (i.e.\ a
context-free grammar generating the text) of size $O(z\log(n/z))$. This
conversion can be performed in $O(z\lg(n/z))$ space and time~\cite{Rytter2003,CLLPPSS05}. Then,
the entire text can be decompressed and streamed in linear time using just the
space of the grammar. 
The problem has also been recently considered in~\cite{belazzougui2016lempel} in the context of external-memory algorithms.
To the best of our knowledge, the only other work addressing small-space LZ77 decompression is~\cite{puglisi2019lempel}, which implements (a practical version of) the ideas described in our paper. In particular, no other theoretical solutions using $O(z)$ space are known.

\subsection{Our contributions}

The main contribution of this paper is to show that LZ77 decompression can be performed in $O(z)$ space and almost linear time (in the length of the extracted string). %Even better, we 
We provide two space-time trade-offs which enable us to achieve either linear time \emph{or} linear space \emph{or} both if the alphabet's size is constant. The first trade-off is particularly appealing on small alphabets, while the second dominates the first on large alphabets and the folklore algorithm based on grammars.

Our solution even works for decompressing any specified subsequence of $S$
with little overheads on top of the linear running time (in the extracted substring's length) and working space. 
%In this respect, note that at least $\Omega(z)$ time is needed to read the input.
As an application, we show that our
techniques yield improved results for pattern matching problems on LZ77-compressed text.

We formalize the LZ77 decompression problem as follows. The input consists of an LZ77 representation of a text (we use the version where phrases and sources are not allowed to overlap) and a list of text substrings encoded as pairs: $(i_1, j_1), \ldots, (i_s, j_s)$. 
We decompress these substrings and output them (e.g. to a stream or to disk) character-by-character in the order $S[i_1, j_1], \ldots, S[i_s, j_s]$.
Since both the input strings and the output can be streamed (for example, from/to disk) we only count the working space used on top of the input and the output.
Let the quantity $l = \sum_{k=1}^s (j_k - i_k + 1)$ denote the total number of characters to be extracted. Our main results are summarized in the following two theorems. Let $S$ be a string of length $n$ from an alphabet of size $\sigma$ compressed into an LZ77 representation with $z$ phrases.

\begin{theorem}\label{thm:small}
	 For any parameter $0\leq \delta \leq 1$, we can decompress any $s$ substrings of $S$ with total length $l$ in $O(l\lg^\delta \sigma + (s + z)\lg(n/z))$
	time using $O(z\lg^{1 -\delta} \sigma)$ space.
\end{theorem}
\begin{theorem}\label{thm:main}
	For any parameter $1 \leq \tau \leq \lg (n/z)$, we can decompress any $s$ substrings of $S$ with total length $l$
	in $O\left(\frac{l \lg(n/z)}{\tau} + (s + z)\lg(n/z)\right)$ time using $O(z\lg \tau)$ space.
\end{theorem}

Theorems~\ref{thm:small} and~\ref{thm:main} lead to a series of new and non-trivial bounds on different algorithmic problems on LZ77.
For instance, we provide a smooth time-space trade-off for decompressing the whole $S$ in $O(n\lg^\delta \sigma)$ time using $O(z\lg^{1-\delta}\sigma)$ space
for any constant $0 \leq \delta \leq 1$. 
%This trade-off yields the $O(n)$-time and $O(z)$-space solution on constant-sized alphabets.
By combining Theorem~\ref{thm:main} with $\tau = \lg(n/z)$ with grammars, we furthermore show how to decompress $S$ in $O(n)$ time using $O(z\lg \lg(n/z))$ space.
Both bounds are strict improvements over the previous best complexity of $O(n)$ time and $O(z\lg(n/z))$ space.
See Section \ref{sec:full} and Corollaries \ref{corollary:decompress_S_1} and \ref{corollary:decompress_S_2} for details.

Our results also imply new trade-offs for the pattern matching and approximate pattern matching problems on LZ77-compressed texts.  By showing how our techniques can be combined with existing pattern matching results, in
%our full version [FULL]
Appendix~\ref{app:pattern} 
we show the following:

\begin{theorem}
	Let $S$ be a string of length $n$ compressed into an LZ77 representation $\mathcal{Z}$ with $z$ phrases,
	let $P$ be a pattern of length $m$
	and let $\mathcal A$ be an algorithm that can detect an (approximate) occurrence of $P$ in $S$
	(with at most $k$ errors) given $P$ and $\mathcal{Z}$
	in $t(z, n, m, k)$ time and $s(z,n, m, k)$ space. Then, 
	we can solve the same task in 
	$O(t(z, zm, m, k) + z\lg(n/z))$ time and $O(s(z, zm, m, k) + z)$ space.
	If $\mathcal A$ reports all $\occ$ occurrences
	using $t(z, n, m, k)$ time and $s(z,n, m ,k)$ space,
	then we can report all occurrences in
	$O(t(z, zm, m, k) + z\lg(n/z) + \occ)$ time and $O(s(z, zm, m, k) + z + \occ)$ space.

	\label{lem:black1}
\end{theorem}
\begin{theorem}
	Let $\mathcal A$ be a streaming algorithm that reports all $\occ$ (approximate) occurrence of a pattern $P\in [\sigma]^m$
	(with at most $k$ errors) in a stream of length $n$ in $t(n, m, k)$ time and $s(n, m, k)$ space.
	Then, we  can report all occurrences of $P$ in the LZ77 representation of a string $S\in[\sigma]^n$
	in either:
	\begin{itemize}
		\item[]$\bullet$ $O(t(zm, m, k) + z\lg(n/z))$ time and $O(s(zm, m, k) + z\lg \lg(n/z) + \occ)$ space or
		\item[]$\bullet$ $O(t(zm, m, k) + z\lg(n/z) + zm \lg^\delta \sigma )$ time and $O(s(zm, m, k) + z \lg^{1-\delta} \sigma + \occ)$ space.
	\end{itemize}
	\label{lem:black2}
\end{theorem}

The best known algorithm for detecting if pattern $P$ occurs in a string $S$
given $P$ and $\mathcal{Z}$ 
uses $O(z\lg(n/z) + m)$ time and $O(z\lg n + m)$ space \cite{Gawrychowski2011}.
If we plug this into Theorem \ref{lem:black1} we obtain $O(z\lg(n/z) + m)$ time and $O(z\lg m + m)$ space
thereby reducing the $\lg n$ factor in the space to $\lg m$ without increasing the time.

We also obtain new trade-offs for reporting all approximate occurrences of $P$ with at most $k$ errors.
For example, if we plug in the Landau--Vishkin and Cole--Hariharan \cite{Landau1989,Cole2002} algorithms,
we can solve the problem in $O(z\lg(n/z) + z\min\{mk, k^4 + m\} + \occ)$ time using $O(z + m + \occ)$ space for constant-sized alphabets
or $O(z\lg \lg(n/z) + m + \occ)$ space for general alphabets.
The previous best solution has the same time complexity but uses $O(z\lg n + m + \occ)$ space \cite{GGP15}.
The complete explanation can be found in 
Appendix \ref{app:pattern}.
%the full version [FULL] of this paper.

\subsection{Related work}
While the LZ77 decompression problem has not been studied much in the
literature, the problem of fast LZ77 \emph{compression} in small working space has lately
attracted a lot of research in the field of compressed
computation~\cite{fischer2015approximating,fischer2015lempel,policriti2015fast,policriti2017lz77,PSC2016-14}.

A closely related problem is the \emph{random access problem}, where the aim is
to build a data structure taking space as close as possible to $O(z)$ words
and supporting efficient access queries to single characters.
Existing solutions for the random access problem \cite{Rytter2003,CLLPPSS05,BILLE201866} need $\Omega(z \log(n/z))$ space to
achieve $O(\log(n/z))$ access time.
Because these data structures can be built efficiently they also solve the LZ77 decompression problem considered in this paper.
In particular they can decompress the entire string $S$ given its LZ77 representation
in  $O(n)$ time using $O(z\lg(n/z))$ working space.
%Our results improve this working space: we show how to decompress $S$ in  $O(n)$ time using only $O(z\lg \lg(n/z))$ working space for general alphabets and $O(z)$ working space for constant-sized alphabets.

Random access data structures can also decompress any set of $s$ substrings with total length $l$
in $O(l + s\lg n)$ time.
We provide several new trade-offs for this problem; for instance we can solve it using
$O(l + (z + s)\lg(n/z))$ time and $O(z)$ space for constant-sized alphabets or $O(z\lg \lg(n/z))$ space for general alphabets.

%Entropy-compressed self-indices such as the FM-index also efficiently solve the random access problem~\cite{ferragina2000opportunistic}. However, the FM-index cannot be used to solve the LZ77 decompression problem because it requires $\Omega(n)$ time and space and access to the uncompressed string at construction time. After construction, this data structure achieves space close to the high-order entropy of the text but always requires at least $\Omega(n/\lg n)$ bits of space which can be exponentially more than what is required by the LZ77 representation of the text.

In a recent work~\cite{puglisi2019lempel}, Puglisi and Rossi implemented the ideas described in our paper. They showed that, even if an implementation of our algorithms is not practical due to the underlying mergeable dictionary, several optimizations can be introduced that drastically improve performance. Their optimized implementation led to new relevant space-time tradeoffs on several datasets of practical interest.

\section{Preliminaries}

We assume a standard unit-cost RAM model with
word size $w = \Theta(\lg n)$ and that the input is from an integer alphabet $\Sigma = \{1, 2, \ldots, \sigma\}$ where $\sigma \leq n^{O(1)}$,
and we measure space complexity in words unless otherwise specified.
A string $S$ of length $n = |S|$ is a sequence $S[1]\ldots S[n]$ of $n$ symbols from an alphabet~$\Sigma$
of size $|\Sigma| = \sigma$.
The string $S[i]\dots S[j]$ denoted $S[i,j]$ is called a \textit{substring} of $S$.
Let $\epsilon$ denote the empty string and let $S[i,j] = \epsilon$ when $i > j$.
To ease the notation, let $S[i,j] = S[1, j]$ if $i < 1$ and $S[i,n]$ if $j > n$.
Let $[u]$ be shorthand for the interval $[1;u] = \{ 1, 2, \ldots, u\}$ and $\$$ be a special symbol that never occurs in the input.
A straight-line program (SLP) is an acyclic grammar in Chomsky normal form
where each non-terminal $T$ has exactly one production rule with $T$ as its left-hand side i.e.,
a grammar where each non-terminal production rule expands to two other rules and generates one string only.

%\subsection
\paragraph{Lempel-Ziv 77 Algorithm}

For simplicity of exposition we use the scheme given by
Farach \& Thorup \cite{Farach1998}. Map $\Sigma$ into $[\sigma]$ and assume that $S$ is prefixed by $\Sigma$ in the negative positions, i.e. $S[-c]=c$ for $c\in\Sigma$ and $S[0]=\$\notin \Sigma$.

An \textit{LZ77 representation} \cite{lempel1976complexity,lempel77} of $S$ is a string
$\mathcal{Z}$ of the form $(s_1, l_1)\ldots (s_z, l_z) \in ([-\sigma; n] \times [n])^z$.
Let $u_1 = 1$ and $u_i = u_{i-1} + l_{i-1}$, for $i > 1$.
For $\mathcal{Z}$ to be a valid LZ77 representation of $S$, we require that $s_i + l_i \leq u_i$ and that
$S[u_i, u_i + l_i - 1] = S[s_i, s_i + l_i - 1]$ for $i \in [z]$.
This guarantees that $\mathcal{Z}$ \textit{represents} $S$ and clearly $S$ is
uniquely defined in terms of $\mathcal{Z}$.
We refer to the substring $S[u_i, u_i + l_i - 1]$ as the \emph{$i^{th}$ phrase}
of the representation, the substring $S[s_i, s_i + l_i - 1]$ as the \emph{source of the
	$i^{th}$ phrase} and $(s_i, l_i)$ as the \emph{$i^{th}$ member} of $\mathcal{Z}$.
%Let $\mathcal{U} = \{u_1, \ldots u_z\}$ be the set of phrase beginnings.
We note that the restriction $s_i + l_i \leq u_i$ for all $i$
implies that a source and a phrase cannot overlap and thus
we do not handle representations that are self-referential.

By the given definition, the LZ77 representation of a string
is not unique, however a minimal $LZ77$ representing a text $S$ can be
found greedily in $O(n)$ time \cite{crochemore2008simple,karkkainen2013linear}.

%\subsection
\paragraph{Mergeable Dictionary}
\label{sec:merge}

To obtain our results we need mergeable dictionaries with shift operations.
The Mergeable Dictionary problem is to maintain a dynamic collection $\mathcal{G}$ of sets $\{G_1, G_2, \ldots \}$
of $n$ elements from an ordered universe $\{1, 2, \ldots, \mathcal{U}\}$ starting from $n$ singleton sets under the operations:

\begin{enumerate}
	\item $C \leftarrow \dmerge{A, B}$: Remove $A$ and $B$ from $\mathcal{G}$ and insert $C = A \cup B$ instead.
	\item $(A, B) \leftarrow \dsplit{G, x}$: Split $G$ into two sets $A = \{ y \in G \mid y \leq x\}$ and
	$B = \{ y \in G \mid y > x \}$. $G$ is removed from $\mathcal{G}$ while $A$ and $B$ are inserted.
	\item $G' \leftarrow \dshift{G,x}$ for some $x$ such that $y + x \in [\mathcal{U}]$ for each $y \in G$:
	Create the set $G' = \{ y + x \mid y \in G\}$. $G$ is removed from $\mathcal{G}$ while $G'$ is inserted.
	\item $\makeset{j}$: Insert a new singleton set $G = \{j\}$ in $\mathcal{G}$.
	
\end{enumerate}

Bille et al.~\cite{Bille2019} show how to extend the mergeable dictionary by Iacono \& \"{O}zkan \cite{Iacono2010} to support shifts (Iacono \& \"{O}zkan \cite{Iacono2010} write that their data structure can be extended to support the shift operation but do not provide any details):

\begin{theorem} [\cite{Bille2019}]
	\label{thm:merge}
	There exists a mergeable dictionary with shifts supporting all operations in $\lg \mathcal{U}$ amortized time using linear space.
	For a set $G$, let $\mathcal{U}_G = \max(G) - \min(G)$. The split operations take $O(\lg \mathcal{U}_G)$ worst-case and amortized time
	and the makeset and shift operations take $O(1)$ worst-case and amortized time. The amortized time of the merge operation is $O(\lg \mathcal{U}_G)$,
	where $G$ is the set output by the operation.
\end{theorem}

\section{LZ77 Induced Context}

In this section we present the centerpiece of our algorithm.
It builds on the fundamental property of LZ77 compression that any substring
of a phrase also occurs in the source of that phrase.
Our technique is to store a short substring, which we call \emph{context}, extracted around phrase borders.
The contexts are stored in a compressed form that allows faster substring extraction 
than that of LZ77. 
We then take advantage of this property when extracting a substring of $S$
by splitting it into short chunks which in turn are extracted by repeatedly mapping them to
the source of the phrase they are part of.
Eventually, they will end up as substrings of the contexts from where they can be efficiently extracted.

The technique resembles what Farach \& Thorup refer to as \textit{winding} in \cite{Farach1998}.
We show new applications of the technique and obtain
better time complexity by using the mergeable dictionary of Section \ref{sec:merge}.

%In Section \ref{sec:context-parse} we show how to obtain an LZ77 parse for the subsequence of $S$ that includes only the context of every phrase. This representation can be efficiently transformed into an SLP as shown in Section \ref{sec:context-storing} using the online construction algorithm by Charikar et al.~\cite{CLLPPSS05} or Rytter \cite{Rytter2003}.

Recall that we assume $S$ is prefixed by the alphabet in the negative positions, that $S[0] = \$$, and
that $u_k$ is the starting position in $S$ of the $k^{th}$ phrase.

\begin{definition}
	Let $\tau$ be a positive integer. The \emph{$\tau$-context} of a string $S$ (induced by an LZ77 representation $\mathcal{Z}$ of $S$)
	is the set of positions $j$ where either $j \leq 0$ or there is some $k$ such that $u_k - \tau < j < u_k + \tau$.
	If positions $i$ through $j$ are in the  $\tau$-context of $S$, then we simply say
	``$S[i,j]$ is in the $\tau$-context of $S$''.
\end{definition}

\begin{figure}[t!]
\begin{center}
	\includegraphics[width=.9\textwidth]{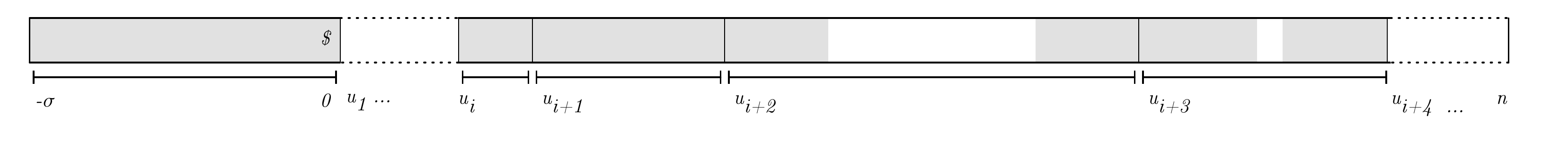}
	\end{center}
	\vspace{-5ex}
	\caption{Example of the $\tau$-context of a string. Dashed parts are truncated parts of the string not shown
		by the figure, grey parts represent substrings in the $\tau$-context and white parts represent
		substrings not in the $\tau$-context. The first substring in the negative positions $-\sigma$ through $0$
		is always in the $\tau$-context. Recall that $l_i$ is the length of the $i^{th}$ phrase. In this example, $l_i < \tau, l_{i+1} \leq 2\tau$ and $l_{i+2}, l_{i+3} > 2\tau$.
	}
\end{figure}

\begin{definition}
	Let $\tau$ be a positive integer. The \emph{$\tau$-context string} of $S$, denoted $S^\tau$, is the subsequence of $S$ that includes $S[j]$
	if and only if $j$ is in the $\tau$-context of $S$. We denote with $\pi^\tau(j)$ the unique position in $S^\tau$ where such a position $j$ is mapped to (i.e. $S^\tau[\pi^\tau(j)] = S[j]$). 
	\label{def:stau}
\end{definition}

It is easy to show how to map positions from $S$ to  $S^\tau$ 
(see Appendix \ref{app:lemma:pi} for a proof)
%(see [FULL] for a proof):

\begin{lemma}\label{lemma:pi}
	Let $\mathcal{Z}$ be an LZ77 representation of a string $S$ of length $n$ with $z$ phrases
	and let $\tau$ be a positive integer.
	Given $t=O(z)$ sorted positions, $p_1 \leq \ldots \leq p_t \in [n]$ in the $\tau$-context of $S$
	we can compute $\pi^\tau(p_1), \ldots, \pi^\tau(p_t)$ in $O(z)$ time and space.
	\label{lem:map}
\end{lemma}

We use $\pi$ as shorthand for $\pi^\tau$ whenever $\tau$ is clear from context.
The following properties follow from the definitions and Lemma~\ref{lem:map}
but will come in handy later on:

\begin{property}
	If $a, a'$ are positions in the $\tau$-context of $S$ and $a < a'$
	then $\pi(a) < \pi(a')$.
	\label{prop:order}
\end{property}
\begin{property}
	If $S[a, b]$ is in the $\tau$-context of $S$
	then  $S^\tau[\pi(a), \pi(b)] = S[a, b]$
	\label{prop:equality}
\end{property}

We now consider the following problem:
given a substring $S[i,j]$ of length at most $\tau$, find a pair of integers $(i', j')$
such that $i' \leq i$, $S[i,j] = S[i', j']$ and $S[i',j']$ is in the $\tau$-context of $S$.

We first give an informal overview of how the algorithm works.
Recall that if a substring of $S$ is contained within a phrase in the LZ77 parse of $S$,
then the substring also occurs in the source of that phrase.
The idea is to repeat this process of finding an identical substring in the source
until the found string is in the $\tau$-context of $S$, which happens after at most $z$ steps. 
To do this efficiently for multiple strings, we use the mergeable dictionary structure to maintain the relevant positions.
This allows us to process all strings inside a phrase simultaneously because they all need to
be moved to the same source. By processing the phrases in right-to-left order we can bound
the number of dictionary operations by the number of phrases.

The following algorithm gives the details of how to solve the
problem for a set of $z$ substrings using $O(z)$ space and $O(z\lg n)$ time (which we later improve).

\begin{algorithm}
	\label{alg:mapping}
	
	Let $\mathcal{Z}$ be an LZ77 representation of a string $S$ of length $n$ with $z$ phrases and let $\tau$ be a positive integer.
	The input is $t = O(z)$ substrings of $S$ given as pairs of integers denoting start and end positions:
	$(a_1,b_1), \ldots, (a_t, b_t)$ where $b_i-a_i < \tau$ for all $i \in [t]$.
	Let $\mathcal{G}$ be a mergeable dictionary as given by Lemma~\ref{thm:merge}.
	For each of the pairs $(a_i,b_i)$ create a singleton set $G_i$ with element $x_i$ at position $a_i$ and finally merge all these
	elements into a single set $G$. Each element $x_i$ has associated its rank $i$ among the input pairs as satellite information.
	
	We now consider the members of $\mathcal{Z}$ one by one in reverse order.
	Member $(s_i, l_i)$ is processed as follows:
	%
	%Assume that $K \in D$ is a special set and that all elements in $K$ are at most $u_i + l_i - 1$. Execute the following operations on $D$:
	\begin{enumerate}
		\item[]1. If $l_i \leq \tau$ skip to the next member.
		\item[]2. Otherwise let
		\begin{enumerate} 
			\item $(A, B) \leftarrow \dsplit{G, u_i + l_i - \tau}$
			\item $(A', B') \leftarrow \dsplit{A, u_i - 1}$,
			\item $B'' \leftarrow \dshift{B', s_i-u_i  }$
			\item $G \leftarrow \dmerge{A', B''}$.
		\end{enumerate}
	\end{enumerate}
	
	In step 1, we skip a phrase if it is no longer than $\tau$ because any string of length $\tau$ or shorter
	starting in that phrase is already in the $\tau$-context of $S$.
	In step 2a, we split the set such that all strings that start in the last $\tau$ positions of the phrase
	are not shifted, because these already are in the $\tau$-context of $S$.
	In 2a-d we split the set to obtain the set of strings $B''$ that starts in the $i^{th}$ phrase
	excluding those starting in the last $\tau$ positions, as they are already in the $\tau$-context
	of $S$. These strings are then shifted to the source and will be considered again in later iterations.
	
	\begin{figure}[t!]
	\begin{center}
		\includegraphics[width=.9\textwidth]{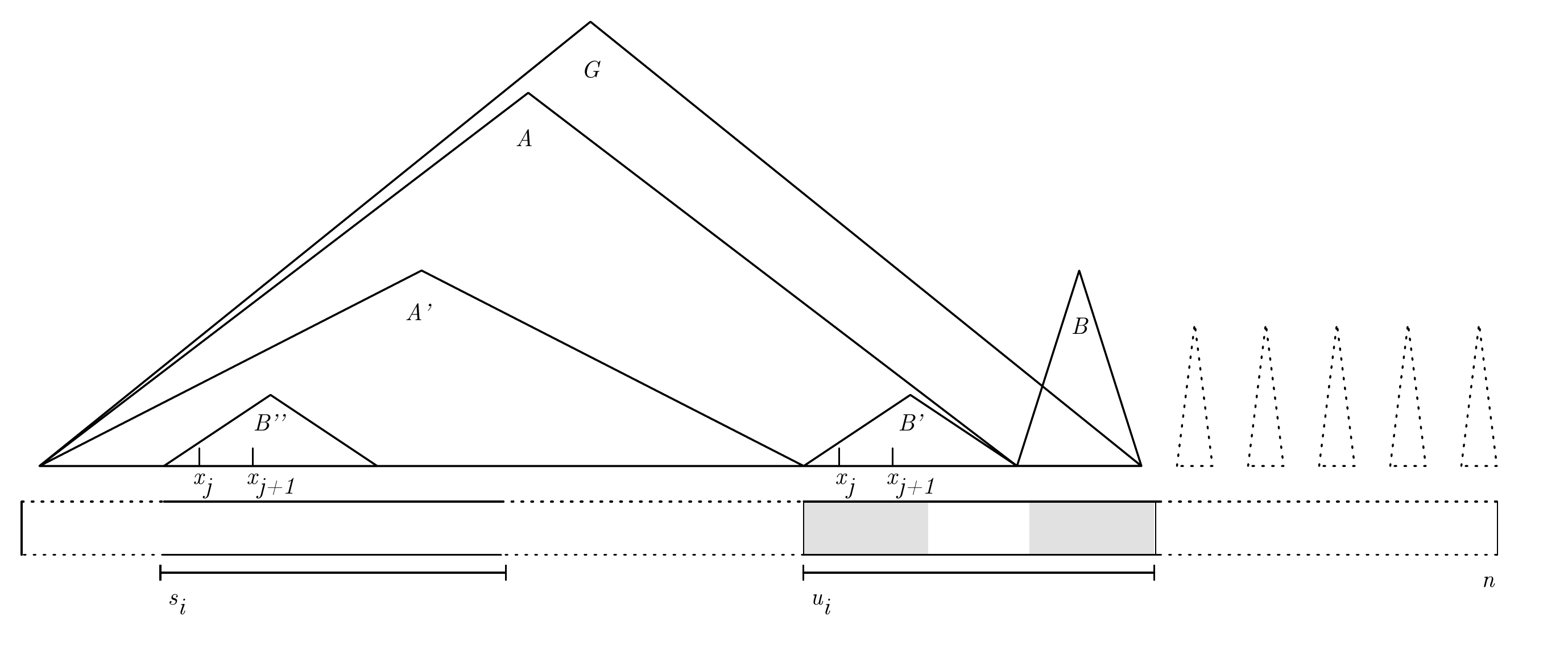}
    \end{center}
    \vspace{-5ex}
		\caption{Example of the dictionaries created during an iteration. The dashed parts of the string are truncated 
			parts not relevant to the example. The dotted triangles represent the $B$-sets from earlier iterations.
			The grey parts show the $\tau$-context of the string inside the $i^{th}$ and $i-1^{th}$ phrase.
			Note that the set $B''$ is the set $B'$ after the shift operation. As exemplified by the elements $x_j$ and $x_{j+1}$, the relative order and position inside the set is unaffected by the shift.
			Let $p$ and $p'$ be the position of $x_j$ before and after the shift, respectively.
			Observe also that $S[p,p+(b_j-a_j)-1] = S[p',p'+(b_j-a_j)-1]$, so the shift does not affect the substring represented by $x_j$.
			An iteration starts from the set $G$, obtaining $A$ by cutting off $B$.
			The new $G$ (not shown in the figure) is then obtained by shifting all elements in the range of $B'$, which are all
			contained in the $i^{th}$ phrase to the same relative position in the source of the phrase.
		}
	\end{figure}
	
	After processing all members, scan each set in $\mathcal{G}$ to retrieve all the elements. Let $p(x_i)$ denote the new position in $\mathcal{G}$ of element $x_i$.
	We then output the pairs $(p(x_1), e_1), \ldots, (p(x_t), e_t)$ in order of their rank $i$ where $e_i = p(x_i) + b_i - a_i$.
	\end{algorithm}

The proof of correctness and the analysis of the time complexity can be found in Appendix \ref{app:mapping}.
%For space constraints, we leave the proof of correctness and the analysis of the time complexity to our full version [FULL].

\paragraph*{Bucketing}

To decrease the running time from $O(z\lg n)$ to $O(z\lg (n/z))$, we apply a bucketing argument similar to the one given by Farach \& Thorup \cite{Farach1998}. The overall idea, described in detail in 
%our full version [FULL], 
Appendix \ref{app:mapping},
is to divide the universe of text positions $[n]$ into $z$ buckets of size $\lceil n/z \rceil$ each, and keep a separate mergeable dictionary for each of the buckets. With a constant additional number of dictionary operations per phrase, we are then able to simulate the dictionary used by Algorithm~\ref{alg:mapping}. Since the universe size of each dictionary is now reduced to $O(n/z)$, dictionary operations cost just $O(\lg(n/z))$ time (amortized) and we obtain:

\begin{lemma}\label{lemma: substrings to tau context}
	Let $\mathcal{Z}$ be an LZ77 representation of a string $S$ of length $n$ with $z$ phrases
	and let $\tau$ be a positive integer.
	Given $t\in O(z)$ substrings of $S$ as pairs of integers $(a_1,b_1), \ldots, (a_t,b_t)$ where $b_i-a_i < \tau$
	we can find $t$ pairs of integers $(a'_1,b'_1), \ldots, (a'_t,b'_t)$ such that $S[a'_i, b'_i] = S[a_i,b_i]$, $a'_i \leq a_i$
	and $S[a'_i, b'_i]$ is in the $\tau$-context of $S$ using $O(z)$ space and $O(z\lg(n/z))$ time.
	\label{lem:wind}
\end{lemma}

\subsection{LZ77 Compressed Context}
\label{sec:context-parse}

It is possible to obtain an LZ77 representation $\mathcal{Z}^\tau$ of the string $S^{\tau}$
directly from an LZ77 representation $\mathcal{Z}$ of $S$. Informally, the idea is to split every phrase of $\mathcal{Z}$
into two new phrases consisting of respectively the first and last $O(\tau)$
characters of the phrase.
In order to find a source for these phrases, we use Algorithm~\ref{alg:mapping}
which finds an identical string that also occurs in $S^\tau$.

We now describe the algorithm sketched above that constructs an LZ77 representation $\mathcal{Z^\tau}$ of $S^\tau$
given the LZ77 parse $\mathcal{Z}$ of $S$.

\begin{algorithm}
	\label{alg:context-parse}
	First we construct $O(z)$ \emph{relevant pairs} of integers representing
	substrings of $S$ by considering the members of $\mathcal{Z}$ one by one in order.
	Member $(s_i, l_i)$ is processed as follows:
	
	\begin{enumerate}
		\item[]1. If $l_i \leq \tau$: Let $(u_i, u_i + l_i - 1)$ be a relevant pair.
		\item[]2. If $\tau < l_i < 2\tau$: Let $(u_i, u_i + \tau - 1)$ and $(u_i + \tau, u_i + l_i - 1)$
		be relevant pairs.
		\item[]3. Otherwise $l_i \geq 2\tau$:
		Let $(u_i, u_i + \tau - 1)$ and $(u_i + l_i - \tau + 1, u_i + l_i - 1)$ be relevant pairs.
	\end{enumerate}
	
	Each of the relevant pairs represents a prefix or a suffix of a phrase. 
	The concatenation of these phrase prefixes and suffixes in left-to-right order
	is exactly the string $S^\tau$.
	Let $(a,b)$ be a relevant pair created when considering the $i^{th}$ member
	of $\mathcal{Z}$. Then we say that $(a', b') = (a - u_i + s_i, b - u_i + s_i)$ is the \emph{related source pair}
	and clearly $S[a,b] = S[a', b']$.
	
	Note that the related source pairs might not be in the $\tau$-context.
	We now use Algorithm~\ref{alg:mapping} 
	to find a pair of integers $(a'', b'')$ for each related source pair $(a',b')$ such that
	$S[a'', b''] = S[a',b']$, $a'' \leq a'$ and $S[a'', b'']$ is in the $\tau$-context of $S$.
	We give the pairs in order of creation and this order is preserved by Algorithm~\ref{alg:mapping}.
	If $(a'',b'')$ is the $i^{th}$ output of Algorithm~\ref{alg:mapping}
	then $(\pi^\tau(a''), l)$ is the $i^{th}$ member of $Z^\tau$ where $l = b-a + 1$ and  $\pi^\tau(a'')$ is computed using Lemma \ref{lemma:pi}.
\end{algorithm}

%Again due to space constraints, we leave the proof of correctness and the analysis of the time complexity to our full version [FULL].
We leave the proof of correctness and the analysis of the time complexity to Appendix~\ref{app:context-parse}.  
We obtain the following lemma:

\begin{lemma}
	Let $\mathcal{Z}$ be an LZ77 representation of a string $S$ of length $n$ with $z$ phrases.
	We can construct an LZ77 representation $\mathcal{Z^\tau}$ of $S^\tau$ with $O(z)$ phrases in $O(z\lg (n/z))$ time and $O(z)$ space.
	\label{lem:context-parse}
\end{lemma}

\subsection{Packed and SLP-Compressed  Context}
\label{sec:context-storing}

In this section we consider how to store in two different (packed/compressed) representations a $\tau$-context string of $S$. Our representations can be built quickly and support fast random access.

Our first solution uses word packing. First, we construct the LZ77 representation of $S^\tau$ using Algorithm~\ref{alg:context-parse} and decompress it naively.
Constructing the representation takes time $O(z \lg(n/z))$
while decompressing it takes linear time in its length, $O(z\tau)$.
A string of length $z\tau$ can be stored in $O(z\tau\lg \sigma/ \lg n)$ words using word packing. 
%We obtain:

\begin{lemma}
	Let $S$ be a string $S$ of length $n$ from an alphabet of size $\sigma$ 
	compressed into an LZ77 representation with $z$ phrases, and let $\tau$ be a positive integer.
	We can build and store $S^\tau$ in $O(z(\lg(n/z) + \tau))$ time and $O(z \tau\lg\sigma/\lg n)$ space.
	\label{lem:context}
\end{lemma}

As an alternative solution, we show how to store the context string as an SLP supporting fast random access. The following lemma follows easily from Charikar et al. \cite{CLLPPSS05} and Rytter \cite{Rytter2003} (for a proof, see 
%our full version [FULL]):
Appendix \ref{app:slp}):

%\begin{theorem}[Charikar et al. \cite{CLLPPSS05}, Rytter \cite{Rytter2003}]
%	Let $z$ be the number of phrases in a LZ77 representation of a string $S'$ of length $n'$.
%	We can build a balanced SLP for $S'$ with height $O(\lg n')$ in $O(z\lg(n'/z))$ space and time.
%	\label{thm:partial}
%\end{theorem}

%Using the above theorem, we obtain the following corollary, with the proof left to Appendix~\ref{app:slp}:

\begin{lemma}
	\label{lem:slp}
	Let $S$ be a string of length $n$ compressed into an LZ77 representation with $z$ phrases, and let $\tau$ be a positive integer.
	We can build a balanced SLP of size $O(z\lg \tau)$ for $S^\tau$ in $O(z\lg (n/z))$ time and $O(z\lg \tau)$ space. Furthermore, the SLP supports extraction of any length-$\ell$ substring of $S^\tau$ in $O(\ell + \log\tau)$ time. 
\end{lemma}

\section{LZ77 Decompression}
\label{sec:full}

We now describe how to apply the techniques described in the previous section
to extract arbitrary substrings of $S$.
Let $S$ be a string of length $n$ compressed into an LZ77 representation with $z$ phrases and let $\tau \leq \log(n/z)$ be a positive integer that we will fix later. We show how to extract $s$ substrings of total length $l$.

Split each substring into consecutive blocks of length $\tau$ (except, possibly, the last for each substring), obtaining at most $l/\tau + s$ blocks. Process a batch of $z$ blocks at a time in left-to-right order. There are at most $O(1 + l/(\tau z) + s/z)$ batches, each containing $z$ blocks.
A batch is processed in $O(z\log(n/z))$ time using Lemma \ref{lemma: substrings to tau context}
thereby finding a substring $s'$ in the $\tau$-context of $S$ for every block $s$ in the batch.

Using Lemma~\ref{lem:slp}, we first build the SLP in $O(z\lg(n/z))$ time and $O(z\lg \tau)$ space. After that, the
$z$ substrings in each batch can be extracted in  $O(z\log\tau + z\tau) = O(z\tau)$ time. Summing up, the time to build the SLP (once) and extract and output all batches is $O(z\lg(n/z) + (1 + \frac{l }{\tau z} + s/z) (z \lg(n/z)+z\tau)) = O((s+z)\lg(n/z) + \frac{l\lg(n/z)}{\tau})$. The total space is $O(z\lg \tau)$. This proves Theorem~\ref{thm:main}. 
If we instead use Lemma~\ref{lem:context}, we spend $O(z(\lg(n/z) + \tau))$ time and $O(z \tau\lg\sigma/\lg n)$ space to build $S^\tau$. After that, the $z$ substrings in each batch can be extracted in $O(z\tau)$ time. Summing up, the time to build $S^\tau$ (once) and extract and output all batches is $O(z(\lg(n/z) + \tau) + (1 + \frac{l }{\tau z} + s/z)(z\lg(n/z) + z\tau)) = O((s+z)\lg(n/z) + \frac{l\lg(n/z)}{\tau})$ while the space
becomes $O(z\tau\lg \sigma / \lg n)$ (on top of the input). To prove Theorem~\ref{thm:small}, we fix $\tau = \lg(n/z) / \lg^\delta \sigma$ for any constant $0 \leq \delta \leq 1$.

Theorems~\ref{thm:small} and \ref{thm:main} immediately yield the following two corollaries on the complexity of decompressing the entire string $S$ (notice that $z\lg(n/z) = O(n)$):

\begin{corollary}\label{corollary:decompress_S_1}
	For any parameter $0\leq \delta \leq 1$, we can decompress $S$ in $O(n\lg^\delta \sigma)$ time using $O(z\lg^{1 -\delta} \sigma)$ space.
\end{corollary}

On large alphabets, we can further improve upon this result by plugging $\tau = \log(n/z)$ into Theorem \ref{thm:main}:

\begin{corollary}\label{corollary:decompress_S_2}
	We can decompress $S$ in $O(n)$ time using $O(z\lg\lg (n/z))$ space.
\end{corollary}

%Note that, when aiming at linear running time,  Corollary \ref{corollary:decompress_S_2} is at least as good as Corollary \ref{corollary:decompress_S_1} whenever $\sigma \in \Omega(\lg (n/z))$ (and asymptotically better for $\log\sigma \in \omega(\lg\lg (n/z))$).

%%%%%%%%%%%%%%%%%%%%%%%%%%%%%%%%%
\section{References}
\bibliographystyle{IEEEbib}
\bibliography{refs}

\newpage
\appendix

\section{Proof of Lemma \ref{lemma:pi}}\label{app:lemma:pi}

\begin{proof}
	Let $\text{gap}_k = \max\{0, l_k - 2\tau+1\}$ be the number of positions inside the $k$-th LZ77 phrase that are not in the $\tau$-context of $S$. 
	
	Let $i,k,L$ be three integers initialized as follows: $i=0$, $k=1$, and $L=0$.
	We keep the following two invariants: 
	\begin{itemize}
		\item[](i) if $i>0$, then $k$ is the smallest integer such that $p_i < u_k+l_k$, i.e. position $p_i$ is in the $k$-th phrase, and 
		\item[](ii) $L$ is the number of positions $j<u_k$ such that $j$ is in the $\tau$-context of $S$ (i.e. $L$ is the length of the prefix of $S^\tau$ containing characters from $S[1..u_k-1]$).
	\end{itemize}
	It is clear that (i) and (ii) hold in the beginning of our procedure. We now show how to iterate through the LZ77 phrases and compute the desired output in one pass. 
	
	Assume that we have already computed $\pi^\tau(p_1), \ldots, \pi^\tau(p_i)$ (or none of them if $i=0$). 
	To compute $\pi^\tau(p_{i+1})$, we check whether $u_k\leq p_{i+1} < u_k+l_k$, i.e. whether $p_{i+1}$ is in the $k$-th phrase. If not, we find the phrase containing $p_{i+1}$ as follows.    
	We set $L \leftarrow L + l_k-\text{gap}_k$, $k\leftarrow k+1$ and repeat until we find a value of $k$ that satisfies $u_k\leq p_{i+1} < u_k+l_k$. It is clear that, at each step, $L$ is still the length of the prefix of $S^\tau$ containing characters from $S[1..u_k-1]$ (i.e. invariant (ii) is maintained).
	
	Once such a $k$ is found, we compute $\pi^\tau(p_{i+1})$ simply adding $L$ to the relative position of $p_{i+1}$ inside its phrase, and subtract $\text{gap}_k$ from this quantity if $p_{i+1}$ is within $\tau$ characters from the end of the phrase.  In more detail, if $p_{i+1} < u_k+\tau$, then $\pi^\tau(p_{i+1}) \leftarrow L + 1 + (p_{i+1} - u_k)$.
	Otherwise,  $\pi^\tau(p_{i+1}) \leftarrow L + 1 + (p_{i+1} - u_k) -  \text{gap}_k$.
	The correctness of this computation is guaranteed by the way we defined $L$ in property (ii).
	
	Note that $k$ is again the smallest integer such that $p_{i+1} < u_k+l_k$ (invariant (i)), so we can proceed with the same strategy to compute $\pi^\tau(p_{i+2}), \ldots, \pi^\tau(p_t)$.
	
	Overall, the algorithm runs in $O(z)$ time and space.
\end{proof}

\section{Proof and Analysis of Algorithm~\ref{alg:mapping}}
\label{app:mapping}

	\paragraph*{Correctness}
	Let $p(x_i)$ denote the position in $\mathcal{G}$ of element $x_i$ at any point of the algorithm.
	We now show that for any element $x_i$, we have $S[a_i, b_i] = S[p(x_i), e_i]$ both before and after considering the $j^{th}$ member of $\mathcal{Z}$.
	Initially, $p(x_i) = a_i$ so this is trivially true before the first iteration.
	
	%$p(x_k) = i_k$ and $x$ \textit{represents} the string $S[i_k,j_k]$, i.e. $S[p(x_k), l_k] = S[i_k,j_k]$.
	Assume by induction that this is true before considering member $j$.
	If $p(x_i) > u_j + l_j - \tau$ or $p(x_i) < u_j$, then $p(x_i)$ will not be changed when considering member $i$.
	Otherwise, $u_j \leq p(x_i) \leq u_j + l_j - \tau$, 
	and thus $S[p(x_i), e_i]$ is a substring of $S[u_j, u_j + l_j - 1]$ which 
	also occurs at the same relative position in $S[s_j, s_j + l_j - 1]$.
	Now $x_i$ is shifted such that $p(x_i) \leftarrow p(x_i) - u_j + s_j$ thereby maintaining the relative
	position inside the two identical strings and it follows that $x_i$ still represents $S[a_i,b_i]$
	after considering member $j$ and thus also before considering member $j-1$.
	
	We now show that, for any element $x_i$,  the string $S[p(x_i), e_i]$ is in the $\tau$-context of $S$
	after considering the last member.
	Observe that when considering member $j$, every element positioned in $S[u_j, u_j + l_j - \tau]$
	is shifted to a position less than $u_j$, because $s_j + l_j \leq u_j$ by definition.
	As we are considering the members in reverse order, this means that every element $x_i$
	must end in a position such that either there is some $k$ such that $u_k + l_k - \tau < p(x_i) < u_k + l_k$
	or $p(x_i) < 0$ which concludes the proof of correctness.

	\paragraph*{Complexity} Creating the $z$ singleton elements with positions in the range $[n]$ and merging them to $G$ takes $O(z\lg n)$ time.
	For every member of $\mathcal{Z}$ we do $O(1)$ dictionary operations.
	All positions remain in the range $[-\sigma; n]$ thus this also takes total time $O(z\lg n)$.
	We can easily compute and store $u_1, \ldots, u_z$ in $O(z)$ time and space.
	Outputting the elements $x_i$ in order of their rank $i$ takes linear time as the ranks are consecutive integers thus the total time is $O(z\lg n)$.
	We never store more than the $O(z)$ elements, thus the total space is $O(z)$.

\section{Bucketing}

	Fix $F = \lceil n/z \rceil$ and transform the parse $\mathcal{Z}$ by splitting
	any phrase that covers a position $kF < n$ for $k = 1,2,\ldots$, i.e.\
	replace the element $(s_i, l_i)$ by elements $(s_i, l'_i)$ and $(s_i + l'_i, l_i - l'_i)$
	if $kF \in [u_i, u_i + l_i]$
	where $l'_i = kF - u_i$. Clearly, the parse remains valid and the number of phrases
	is at most doubled \cite{Farach1998}.
	
	We now change the way we merge the sets.
	Initially, create an array $M$ of length $z - 1$.
	Now for each of the pairs $(a_i, b_i)$ we create the singleton set $G_i$
	with element $x_i$ at position $(a_i \mod F)$ and store
	a pointer to $G_i$ in position $\lfloor a_i / F \rfloor$ of $M$.
	If the position is occupied by some set $G_j$ merge the sets $G_i$ and $G_j$ and update
	the position to point at the new set $G_i \cup G_j$. 
	Assume that $\tau \leq n/z$ (this will be true later for the values of $\tau$ we will use).
	When considering the members of $\mathcal{Z}$ in reverse order we process member
	$(s_i, l_i)$ as follows:
	
	\begin{enumerate}
		\item[]1. If $l_i \leq \tau$ skip to the next member.
		\item[]2. Otherwise let
		\begin{enumerate}
			\item $(A, B) \leftarrow \dsplit{M[\lfloor u_i / F \rfloor], ((u_i \mod F) + l_i -\tau)}$.
			\item $(A', B') \leftarrow \dsplit{A, u_i -1}$.
			\item Let $M[\lfloor u_i / F \rfloor]$ point to $A'$.
			\item $B'' \leftarrow \dshift{B', s_i - u_i \mod F}$.
			\item If $s_i - u_i < 0$ continue with the next member.
			\item Otherwise let $G' \leftarrow \dmerge{M[\lfloor (s_i - u_i) / F \rfloor], B''}$.
			\item Let $M[\lfloor (s_i - u_i) / F \rfloor]$ point to $G'$.
		\end{enumerate}
	\end{enumerate}
	%\end{enumerate}
	
	After processing all members, scan each set in $\mathcal{G}$ and proceed as originally described.
	It is fairly easy to verify that this procedure positions all elements identically  
	to the original procedure. Therefore, the correctness follows from above.
	There are still $O(1)$ dictionary operations for each of the $O(z)$ members,
	but the positions of the elements are now in the range $[n/z]$ thus the total time becomes $O(z\lg (n/z))$.
	To retrieve all the elements, let $p(x_i)$ denote the new position in $\mathcal{G}$ of element $x_i$.
	We then output the pairs $(p(x_1), e_1), \ldots, (p(x_t), e_t)$ in order of their rank $i$ where $e_i = p(x_i) + b_i - a_i$.

\section{Proof and Analysis of Algorithm~\ref{alg:context-parse}}
\label{app:context-parse}

	\paragraph{Correctness}
	Let $(a_1,b_1), \ldots, (a_t, b_t)$ be the relevant pairs in order of creation,
	$(a'_1, b'_1) \ldots (a'_t, b'_t)$ be the related source pairs,
	$(a''_1, b''_1) \ldots (a''_t, b''_t)$ be the output of Algorithm~\ref{alg:mapping},
	and let $l_i = b_i - a_i + 1$.

	Our goal  is to show that $\mathcal{Z^\tau} = (\pi(a''_1), l_1), \ldots, (\pi(a''_t), l_t)$ is a valid LZ77 representation of $S^\tau$, that is: (i) the concatenation of the phrases of $\mathcal{Z^\tau}$ yields $S^\tau$, (ii) phrases of $\mathcal{Z^\tau}$ are equal to their sources, and (iii) phrases of $\mathcal{Z^\tau}$ do not overlap their sources. Note that $\mathcal{Z^\tau}$ consists of at most $2z$ phrases
	
	(i-ii) It follows directly from Definition~\ref{def:stau} that the concatenation of the strings
	represented by the relevant pairs in order of creation is $S[a_1, a_1] \cdots S[a_t, b_t] = S^\tau $.
	Since $S[a_i, b_i] = S[a'_i, b'_i]$ and, by Lemma~\ref{lem:wind},
	$S[a''_i,b''_i] = S[a'_i, b'_i]$ then we also have that $S^\tau = S[a''_1, a''_1] \cdots S[a''_t, b''_t]$.
	Now, observe that since $S[a_i, b_i]$ and $S[a''_i, b''_i]$ are in the $\tau$-context of $S$ then by Property~\ref{prop:equality} 
	we have $S[a_i, b_i] = S^\tau[\pi(a_i), \pi(b_i)]$ and $S[a''_i, b''_i] = S^\tau[\pi(a''_i), \pi(b''_i)]$.
	This proves properties (i) and (ii).
	
	(iii) By definition of the LZ77 representation of $S$ and since the substring represented by the pair $(a_i, b_i)$ is entirely contained 
	in a phrase we must have $a'_i + l_i \leq a_i$
	and therefore, by Lemma \ref{lem:wind},  $a''_i \leq a'_i$.
	But this means that $a''_i + l_i \leq a_i$
	and therefore,  by Property~\ref{prop:order},  $\pi(a''_i + l_i) \leq \pi(a_i)$, i.e. property (iii) holds.

	\paragraph{Complexity}
	For every member of $\mathcal{Z}$ we create at most two relevant substrings taking total $O(z)$ time and space.
	Applying Algorithm~\ref{alg:mapping} takes time $O(z\lg (n/z))$ and $O(z)$ space.
	We can easily compute and store $u_1, \ldots, u_z$ by computing $\pi^\tau(a')$ for every substring reported by Algorithm~\ref{alg:mapping}. This takes
	total $O(z)$ time and space using Lemma~\ref{lem:map}, 
	thus the total time is $O(z\lg (n/z))$ and the total space is $O(z)$.

\section{Proof of Lemma~\ref{lem:slp}}
\label{app:slp}

\begin{proof}
	We build the LZ77 representation of $S^\tau$ using Lemma \ref{lem:context-parse} and then convert it into an SLP using the procedure described by Charikar et al. and Rytter:

    \begin{theorem}[Charikar et al. \cite{CLLPPSS05}, Rytter \cite{Rytter2003}]
    Let $z$ be the number of phrases in a LZ77 representation of a string $S'$ of length $n'$.
	We can build a balanced SLP for $S'$ with height $O(\lg n')$ in $O(z\lg(n'/z))$ space and time.
	\end{theorem}
	
	Note that, in our case, $n' = |S^\tau| \leq z\tau$ and the SLP's size  is $O(z\lg(n'/z)) = O(z\lg(z\tau/z)) = O(z\lg\tau)$. To reduce the SLP's height to $O(\log(n'/z)) = O(\log\tau)$, we expand the starting nonterminal until obtaining a sequence of $t=\Theta(z)$ nonterminals $X_1,\dots, X_t$ (which happens at depth $\Theta(\log z)$, the SLP being balanced). 
	We cut those levels from the SLP and add a new starting nonterminal $S' \rightarrow X_1\dots X_t$.
	By navigating the grammar in $O(z\log\tau)$ time, we can associate to each nonterminal the length of the text substring it expands to. This is already sufficient to navigate the grammar (by standard techniques) in $O(\log\tau)$ time starting from one of the symbols of the expansion of $S'$. To get the precise starting symbol, we need one additional predecessor query on the cumulative sizes of the expansions of $X_1, \dots, X_t$, which however can be performed in $O(\log(n'/z)) = O(\log\tau)$ time using, e.g. an Elias-Fano predecessor data structure. To conclude, once reached the leaf (terminal) corresponding to the first character to be extracted, the remaining characters can be retrieved in amortized constant time each by standard techniques, i.e. moving to the next sibling or, if this belongs to a different nonterminal, moving upwards until reaching an ancestor of the next symbol to be extracted, and then descending again to the corresponding leaf. Overall, this procedure visits a sub-tree (of the grammar tree) of size $O(\ell)$.
\end{proof}

\section{Applications in Pattern Matching}
\label{app:pattern}

In this section we show how our techniques can be applied as a black box in combination 
with existing pattern matching results. 

Let $S$ be a string of length $n$ and let $P$ be a pattern of length $m$.
The classical \emph{pattern matching problem} is to report all starting positions of
occurrences of $P$ in $S$.
In the \emph{approximate pattern matching problem} we are given an error threshold $k$ in addition
to $P$ and $S$. The goal is to find all starting positions of substrings
of $S$ that are within distance $k$ of $P$ under some metric, e.g. \emph{edit distance} where the distance is the number
of edit operations required to convert the substring to $P$.
When considering \emph{the compressed pattern matching problem}, the string $S$ is given in some compressed form.
Sometimes, we are only interested in whether or not $P$ occurs in $S$.

Pattern matching on LZ77 compressed texts usually 
takes advantage of the property that any substring of a phrase also occurs
in the source of the phrase.
This means that if an occurrence of $P$ is
contained in single phrase, then there must also
be an occurrence in the source of that phrase.
The implication is that the occurrences of $P$ can be split
into two categories: the ones that overlap two or more phrases and 
the ones that are contained inside a single phrase --- usually referred to as primary
and secondary occurrences, respectively~\cite{Karkkainen95lempel-zivparsing}.
The secondary occurrences can be found from the primary 
in $O(z + \occ)$ time and space \cite{GGP15}
where $z$ is the number of phrases in the LZ77 representation of $S$
and $\occ$ is the total number of occurrence of $P$ in $S$.

\paragraph*{Approximate pattern matching in small space}

Consider the \$-padded $m$-context string of $S$ denoted $S^{\$m}$ obtained by replacing each of the maximal substrings of $S$
that are not in the $m$-context by a single copy of the symbol $\$$. This string has length $O(zm)$.
Observe that all the primary occurrences of $P$ in $S$ are
in this string, that any occurrence of $P$ in this string
corresponds to a unique primary or secondary occurrence of $P$ in $S$
and that we can map these occurrences to their position in $S$ in $O(\occ + z\lg z)$ time and $O(z)$ space
using the same technique as Lemma~\ref{lem:map} but by adding the gap lengths instead of subtracting them.

We now prove Theorem~\ref{lem:black1}.
Let $\mathcal{A}$ be the (approximate) pattern matching algorithm from the theorem
using $t(z,n,m, k)$ time and $s(z,n,m, k)$ space.
The idea is to run algorithm $\mathcal{A}$ on an LZ77 representation of $S^{\$m}$ to find all the primary occurrences.

Our algorithm works as follows. First create an LZ77 representation
$\mathcal{Z}^{\$m}$ of $S^{\$m}$ using
Algorithm~\ref{alg:context-parse}. 
If two consecutive phrases of
$\mathcal{Z}^{\$m}$ are both induced by the same phrase of $\mathcal{Z}$ of length
$2m$ or more we add a phrase between them representing only the symbol $\$$. This
is easy to do as part of Algorithm \ref{alg:context-parse} without changing its complexity
and the result is exactly an LZ77 representation of $S^{\$m}$ with $O(z)$ phrases.
The pattern $P$ occurs in $S$ if and only if it occurs in $S^{\$m}$.
Thus we can run algorithm $\mathcal{A}$ on $\mathcal{Z}^{\$m}$ to detect
an (approximate) occurrence of $P$ in $S$.
Constructing $\mathcal{Z}^{\$m}$ takes $O(z\lg (n/z))$ time and $O(z)$ space
thus the total time becomes $O(z \lg (n/z) + t(z,zm,m, k))$ and the space is $O(z + s(z,zm,m, k))$.

All primary occurrences are found by finding all occurrences of $P$
in $S^{\$m}$, mapping them to their positions in $S$ and filtering
out the secondary occurrences. Hereafter, all the secondary occurrences
can be found in $O(z + \occ )$ time and space \cite{GGP15}.
Mapping and filtering also takes $O(\occ + z)$ time and space.
Thus, if algorithm $\mathcal{A}$ reports all (approximate) occurrences of $P$ in $S$ in $t(z,n,m, k)$ time and $s(z,n,m, k)$ space
we can report all (approximate) occurrences of $P$ in $S$
in $O(z \lg (n/z) + t(z,zm,m, k) + \occ)$ time and $O(z + s(z,zm,m, k) + \occ)$ space.

We now prove Theorem~\ref{lem:black2}.
Let $\mathcal{A}$ be the streaming algorithm from the theorem that reports all (approximate)
occurrences of $P$ in a stream of length $n$ using $t(n,m, k)$ time and $s(n,m, k)$ space.
The idea is to run algorithm $\mathcal{A}$ on the string $S^{\$m}$, which we will stream in chunks to find all the primary occurrences.
We use the same technique as above to first filter the primary occurrences
and then find all the secondary occurrences in $O(z + \occ)$ time and space.

We stream the string $S^{\$m}$ consisting of $O(z)$ substrings of total length $O(zm)$.
We can easily compute when to output a $\$$ during the substring extraction. 
Thus using Theorems \ref{thm:small} or \ref{thm:main} we can stream $S^{\$m}$ in
either $O(zm)$ time using $O(z\lg \lg (n/z))$ space or $O(zm\lg^\delta \sigma + z\lg (n/z))$ time using $O(z\lg^{1-\delta} \sigma)$ space.
The total time is then either:
\begin{itemize}
	\item $O(t(zm, m, k) + z\log (n/z))$ time and $O(s(zm, m, k) + z\lg \lg (n/z) + \occ)$ space or
	\item $O(t(zm, m, k) + z\lg (n/z) + zm\lg^\delta \sigma)$ time and $O(s(zm,m, k) + z\lg^{1-\delta} \sigma + \occ)$ space.
\end{itemize}

\noindent \textbf{Compressed Existence}
Gawrychowski~\cite{Gawrychowski2011} shows how to decide if $P$ occurs in $S$ given $P$ and the LZ77 representation
of $S$ using $O(z\lg(n/z) + m)$ time and space.
Applying Theorem~\ref{lem:black1} we get the following:
\begin{corollary}
	We can detect an occurrence of a pattern $P$ of length $m$ given the LZ77 representation of $S$ 
	in $O(z\lg (n/z) + m)$ time using $O(z\lg m + m)$ space.
	\label{cor:detect}
\end{corollary}

\subsubsection*{Approximate Pattern Matching}
By combining the Landau-Vishkin and Cole-Hariharan \cite{Landau1989,Cole2002} algorithms 
all approximate occurrences with at most $k$ errors on a stream of length $n$
can be found in $O(\min\{nk, nk^4/m + n\})$ time using $O(m)$ space.
Gagie et al. \cite{GGP15} shows how this algorithm can be used to
solve the same problem given an LZ77 representation of $S$
in $O(z\lg n + z\cdot \min\{mk, k^4 + m\} + \occ)$ time and $O(z\lg n + m + \occ)$ space.
Applying Theorem~\ref{lem:black2} to the combined Landau-Vishkin and Cole-Hariharan algorithm we get the following new trade-offs:
\begin{corollary}
	We can report all approximate occurrences of a pattern $P$ of length $m$ with $k$ errors given the LZ77 representation with $z$ phrases
	of a string $S$ of length $n$ in: 
	\begin{itemize}
		\item[]$\bullet$ $O(z\lg (n/z) + z \cdot \min\{mk, k^4 + m\} + \occ)$ time using $O(z\lg \lg (n/z) + m + \occ)$ space or
		\item[]$\bullet$ $O(z\lg (n/z) + z \cdot \min\{mk, k^4 + m\} + zm\lg^\delta \sigma + \occ)$ time using $O(z\lg^{1-\delta} \sigma + m + \occ)$ space.
	\end{itemize}
	\label{cor:pattern}
\end{corollary}

\section{Conclusions}

In this paper we have described the first solution for decompressing Lempel-Ziv 77 in linear time using a space proportional to the input's size on constant alphabets. On general alphabets, we presented a trade-off that allows getting either linear time or linear space. Our solutions can, in general, decompress any subsequence of the text. Our work leaves several open problems. First of all, our solutions for general alphabets cannot achieve both linear time and space. We also note that our running times could be improved by fully exploiting packed computation; while it is definitely possible to slightly improve running times of Theorems \ref{thm:small} and \ref{thm:main} in this sense (at the price of a higher space usage), the (opportunely adjusted) case analysis of Corollaries \ref{corollary:decompress_S_1} and \ref{corollary:decompress_S_2} would not yield optimal packed extraction times. We therefore suspect that a different technique is needed in order to achieve optimality. Finally, for simplicity of exposition we did not consider self-referential LZ77: allowing overlaps represents a further interesting line of extension of our work. 

\end{document}